\documentclass[times, twocolumn]{aastex62}
\usepackage{amsmath,amssymb}
\usepackage{array}
\usepackage{graphicx}
\usepackage{subfigure}
\usepackage{enumitem}
\usepackage{natbib}
\usepackage[normalem]{ulem}
\usepackage{xcolor}

\def\equationautorefname~#1\null{Equation~(#1)\null}

\newcommand{\HI}{\mbox{H\,\textsc{i}}}
\newcommand{\HII}{\mbox{H\,\textsc{ii}}}
\newcommand{\NVII}{\mbox{N\,\textsc{vii}}}

\newcommand{\OVII}{\mbox{O\,\textsc{vii}}}
\newcommand{\OVIII}{\mbox{O\,\textsc{viii}}}
\newcommand{\NeIX}{\mbox{Ne\,\textsc{ix}}}
\newcommand{\NeX}{\mbox{Ne\,\textsc{x}}}

\newcommand{\FeXVII}{\mbox{Fe\,\textsc{xvii}}}
\newcommand{\FeXVIII}{\mbox{Fe\,\textsc{xviii}}}
\newcommand{\FeXIX}{\mbox{Fe\,\textsc{xix}}}
\newcommand{\FeXX}{\mbox{Fe\,\textsc{xx}}}

\newcommand{\chandra}{Chandra}
\newcommand{\xmm}{XMM-Newton}
\newcommand{\rgs}{RGS}
\newcommand{\nustar}{NuSTAR}

\def\kmps{\hbox{km $\rm{s^{-1}}$}}

\newcommand{\scriptname}[1]{\texttt{#1}}
\newcommand{\modelname}[1]{\textit{#1}}

\begin{document}
\title{Spectral modeling of charge exchange in the central region of M51}

\author{Hang Yang}
\affiliation{Purple Mountain Observatory, CAS, Nanjing 210023, P.R.China}
\affiliation{School of Astronomy and Space Science, University of Science and Technology of China, Hefei 230026, P.R.China}

\author{Shuinai Zhang}
\affiliation{Purple Mountain Observatory, CAS, Nanjing 210023, P.R.China}
\affiliation{Key Laboratory of Dark Matter and Space Astronomy, CAS, Nanjing 210023, People's Republic of China}

\author{Li Ji}
\affiliation{Purple Mountain Observatory, CAS, Nanjing 210023, P.R.China}
\affiliation{Key Laboratory of Dark Matter and Space Astronomy, CAS, Nanjing 210023, People's Republic of China}

\correspondingauthor{Shuinai Zhang}
\email{snzhang@pmo.ac.cn}

\begin{abstract}
Charge exchange (CX) emission reveals the significant interaction between neutral and ionized interstellar medium (ISM) components of the dense, multiphase, circumnuclear region of a galaxy.
We use a model including a thermal and a CX components to describe the high-resolution XMM-Newton/RGS spectrum of the diffuse emission in the central region of M51. 
Representative signatures of CX emission -- especially the prominent \OVII\ forbidden line and the excess emission in the \OVIII\ Ly$\gamma$ lines -- can be well explained by the model.
Combined with the \chandra\ images in the \OVIII\ and the \OVII\ bands, we find the soft X-ray emission is dominated by the jet-driven outflow and its interaction with the ambient neutral material.
The jet-driven outflow itself is likely a thermal plasma of $\sim 0.59$~keV, with mostly sub-solar abundances.
It runs into the ambient neutral gas, and produces significant CX emission that accounts for one-fifth of the diffuse X-ray emission in the 7--28~\AA\ band.
The effective interface area in the CX process is one order of magnitude greater than the geometrical surface area of the jet-driven outflow.
The tenuous outflow driven by the nuclear star formation may also contribute a small portion to both the diffuse thermal and CX emission. 
The photoionization by the active galactic nuclei (AGNs) and the resonance scattering by the hot gas itself are disfavored, though the effects from past AGN events may not be ruled out.
\end{abstract}

\section{INTRODUCTION}
Circumnuclear environments in galaxies where the interstellar medium (ISM) is involved in various processes and transported along different paths are a subject of great importance and interest.
The circumnuclear ISM consists of materials both externally acquired and from local stellar ejecta, not only feeding the central engine, but also in the form of feedback due to the nuclear star formation (SF) and supermassive black hole (SMBH) activity.
The stellar feedback in the form of strong stellar winds from young OB stars or explosions of supernovae can power nuclear outflow \citep{veilleux2005galactic, creasey2013supernova}.
Active galactic nucleus (AGN) winds and jets can shock the ambient ISM and induce large-scale outflows \citep{faucher2012physics, heckman2014coevolution}.
These sorts of feedback can effectively affect the circumnuclear environment \citep[e.g.,][]{dimatteo2005energy, nayakshin2009competitive,  kormendy2013coevolution}, and hence influence the host galaxies \citep[e.g.,][]{ferrarese2000fundamental, gebhardt2000relationship, mclaughlin2006relation}. 
Understanding of the life cycle of the circumnuclear ISM is essential for the study of galaxy evolution \citep[e.g.,][]{li2009m31, lena2015complex, schartmann2018life}.

Inevitably, the ISM will contain multiphase gas, that is, both in neutral and ionized forms.
Since the timescales of both dynamical and thermal processes are relatively short in circumnuclear regions, the various ISM components are easily led to interactions.
The details of these interactions, such as the velocity, interface area, and changes of dynamics and energetics, are not well constrained by observations and are not currently fully understood.
Moreover, the charge exchange (CX) emission process, where highly ionized elements capture electrons from neutral atoms or molecules and subsequently de-excite, may be non-negligible for such interacting, multiphase gas  \citep{lallement2004contribution}.
As a result, the CX emission supplies a valuable tool for studying the multiphase gas interactions.

The CX has been investigated in the X-ray regime for some nearby galaxies.
A diffuse X-ray-emitting component is ubiquitous in the nuclear region of nearby galaxies \citep{ho2008nuclear}, and is normally believed to be thermal plasma at collisional ionization equilibrium (CIE) state \citep[e.g.,][]{page2003xray, starling2005xray}. 
However, on the basis of diagnostics using the \OVII\ He$\alpha$ triplet-line ratio, \citet{liu2012cxe} tested the significance of CX in stellar-driven outflows, and found that some star-forming galaxies show line ratios consistent with CX.
In the case of the prototype starburst galaxy M82, the CX contributes about a quarter of the diffuse soft X-ray flux, and the interface area between the neutral and ionized gas is about one order of magnitude larger than the geometry surface of its outflow \citep{zhang2014m82}.

Here we present a case study of the circumnuclear region of the grand design spiral galaxy M51 (also known as the Whirlpool galaxy; $d\sim$8.58~Mpc, \citealt{mcquinn2016distance}).
Unlike the edge-on galaxy M82, where the inner nuclear region is blocked by the disk, more emission from the circumnuclear ISM can be observed in M51, because it is a nearly face-on galaxy \citep[inclination angle: $\sim 20^\circ$;][]{hu2013new}.
M51 is considered as a ``quiescently star-forming galaxy,'' with a mean SF rate density $\sim 0.015~M_\odot$~yr$^{-1}$~kpc$^{-2}$ \citep{calzetti2005star}, although this value can be an order of magnitude higher in its central region \citep{leroy2017cloudscale}.
On the other hand, M51 possesses an obscured low-luminosity AGN \citep{fukazawa2001excess} of intrinsic bolometric luminosity $L_\mathrm{bol} \sim 10^{41}$~erg~s$^{-1}$ \citep{xu2016fluorescent, brightman2018long}.
The existence of AGN radio jets affects the ambient ISM and produces radio lobes \citep{crane1992radio, bradley2004physical}. 
Structures similar to the radio lobes are also detected in the X-ray regime, indicating that the ISM is shock-heated to the ionized state \citep{terashima2001chandra}.
In addition, both galactic molecular inflows \citep{querejeta2016gravitational} and outflows \citep{querejeta2016agn} are reported in M51, making the circumnuclear environment in M51 an ideal laboratory for the study of the interaction of the ISM.

In this work, we study the circumnuclear region in M51 through X-ray observations.
The strong \OVII\ forbidden line has been reported in its X-ray spectrum, suggesting the occurrence of CX in the interface between neutral and ionized media \citep{liu2012cxe, liu2015diffuse}.
Thanks to the high spectral resolution of the \xmm /\rgs\ data, we can seek the signatures of CX emission and fit the entire spectrum using the newest CX model\footnote{http://www.atomdb.org/CX/} \modelname{ACX2} \citep{smith2012approximating}.
Combined with the high spatial resolution of the \chandra\ images, we can further study the interaction among the various components of the circumnuclear ISM.

This paper is structured as follows. We describe the used observational data and data reduction in \autoref{sect:data}. In \autoref{sect:analysis}, we perform a preliminary analysis of the spectral features, the results of which will pave the ground for the spectral analysis detailed in \autoref{sect:fitting}. In \autoref{sect:discussion}, we discuss interpretations and implications of our results, and other possible scenarios. In \autoref{sect:summary} we give a brief summary of our findings. Throughout the paper, errors refer to the 90\% confidence level if not explicitly specified.

\section{DATA REDUCTION}
\label{sect:data}
\subsection{\xmm /\rgs\ Spectra}
We make use of nine archival \xmm /\rgs\ observations covering M51, as listed in \autoref{tab:obsid}.
The data reduction process follows the standard procedure provided by the Science Analysis System (SAS; version:~15.0.0), which removes the flare periods and in turn generates the event files. 
Then spectra of each observation are extracted with respect to the bulge region of M51 (white circle in \autoref{fig:obsreg}). 
However, for grating spectroscopy, it is only possible to confine the spatial regions of light in the cross-dispersion direction, which are denoted as parallel arrows in \autoref{fig:obsreg}. 
The spatial width of all the extraction regions is $\sim 1'$ (2.50 kpc) for observations before 2011, corresponding to the diameter of M51's bulge.
For recent public observations (after 2018), since the pointings are away from the center of M51, only part of the bulge region ($0.6'$) can be included, which still contains most of the emission.
Specifically, the spectra are extracted with respect to the X-ray-brightest position of M51 (R.A.:13$^\mathrm{h}$29$^\mathrm{m}$52.814$^\mathrm{s}$ decl.:+47$^\circ$11$^\prime$39.83$^\prime\prime$ or R.A.:202.47006$^\circ$ decl.:47.194397$^\circ$), which is slightly offset from the optical center of M51. 
Moreover, the background spectra of each observation are modeled by the \scriptname{rgsproc} script automatically. 
At last, all the extracted spectra, both \rgs 1 and \rgs 2 spectra included, are combined together to generate a single spectrum, by using the \scriptname{rgscombine} script. 
The total effective exposure time is about 373~ks.\par 

\begin{deluxetable}{ccccc}[htb!]
\tablecaption{\xmm /\rgs\ observations of M51 \label{tab:obsid}}
\tablecolumns{5}
\tablewidth{0pt}
\tablehead{
\colhead{ObsID} & \colhead{Date} & \colhead{$t_{\rm obs}$} & \colhead{$t_{\textrm{eff}}$} & \colhead{PA}\\
\colhead{} & \colhead{(\textit{Y-M-D})} & \colhead{(ks)} & \colhead{(ks)} & \colhead{($^{\circ}$)}
}
\startdata
0212480801 & 2005--07--01 & 49.214 & 17.961 & 294.6\\
0303420101 & 2006--05--20 & 54.114 & 30.928 & 326.6\\
0303420201 & 2006--05--24 & 36.809 & 22.923 & 323.4\\
0677980701 & 2011--06--07 & 13.319 & 13.205 & 312.3\\
0677980801 & 2011--06--11 & 13.317 & 5.877  & 326.8\\
0824450901 & 2018--05--13 & 77.000 & 76.685 & 325.9\\
0830191401 & 2018--05--26 & 98.000 & 82.438 & 325.8\\
0830191501 & 2018--06--13 & 63.000 & 61.705 & 307.9\\
0830191601 & 2018--06--15 & 63.000 & 61.708 & 306.4\\
\enddata
\tablecomments{The symbol $t_{\rm obs}$ denotes the duration of each observation, and $t_{\textrm{eff}}$ is the effective exposure time after filtering the flare periods. PA is the abbreviation of position angle.}
\end{deluxetable}

\subsection{\chandra\ Image}
Ten \chandra\ observations (ObsID: 13812, 13813, 13814, 13815, 13816, 15496, 15553, 1622, 354, and 3932) are used to produce a stacked image of M51. The total effective exposure time after removing the flare period time is $\sim 835$~ks. The reprocessing is carried out by using the Chandra Interactive Analysis of Observations (CIAO; version:~4.8) scripts. Each observation is reprocessed with \scriptname{\detokenize{chandra_repro}}, which recalibrates the original data and generates level2 event files. 
The final merged image is in the 0.5-1.2 keV band, comparable to the wavelength range of our \rgs\ spectrum, and will be used in spectral fitting in order to account for the effect of spatial broadening (see \autoref{sect:prof}).

\begin{figure}[htb!]\centering
  \includegraphics[width=0.48\textwidth]{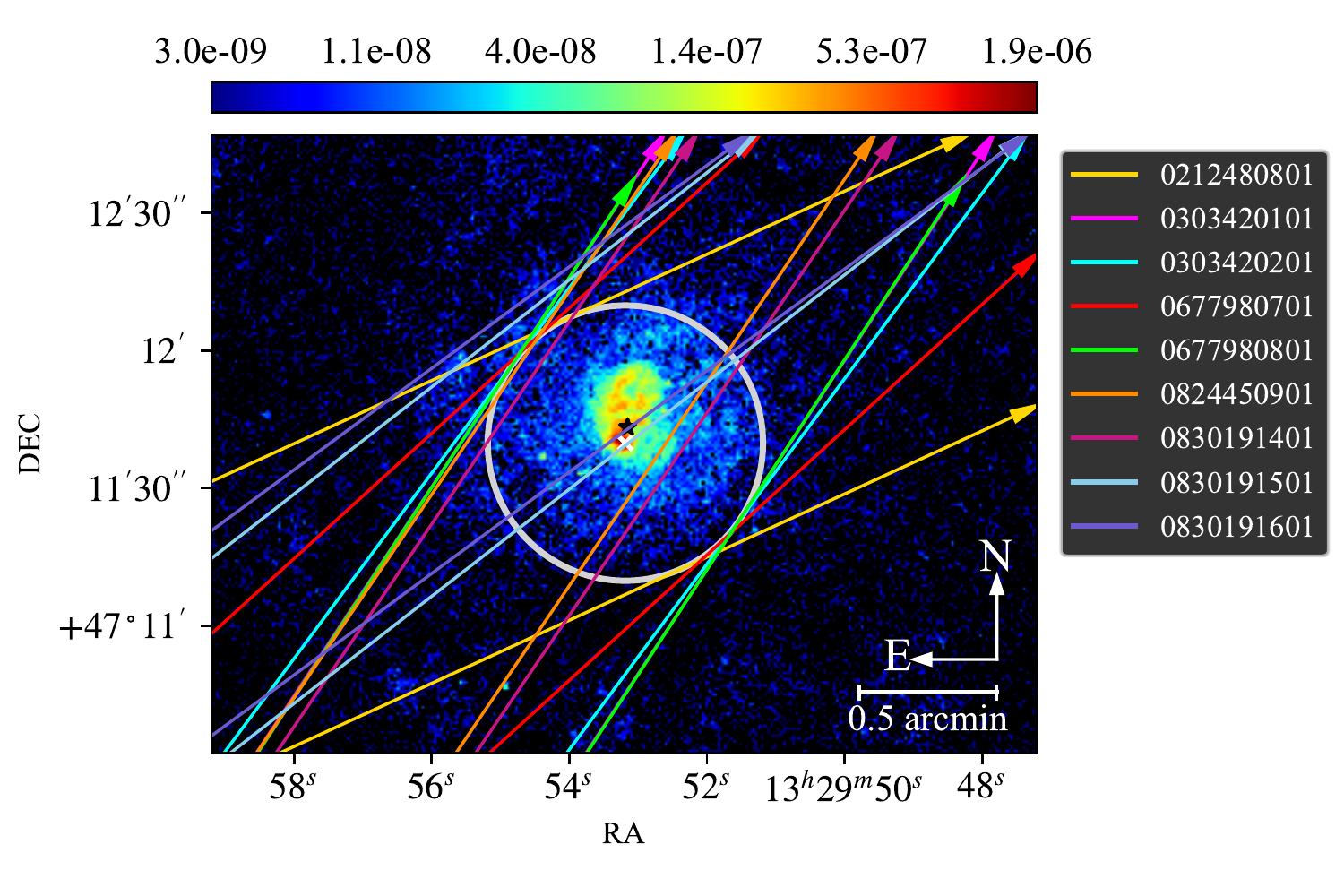}
\caption{\chandra\ image (0.5--1.2~keV) of diffuse emission of M51 in units of photons~cm$^{-2}$~s$^{-1}$, overlaid with the pointing and dispersion direction of each \xmm\ observation. 
The X-ray brightest spot is labeled with a white ``x'' marker; centered at it with a diameter of $1^\prime$, the white circle represents the circumnuclear region of M51. 
The black star is the position of the SMBH. 
Each observation listed in \autoref{tab:obsid} is plotted in a distinctive color.
The parallel arrows with corresponding colors denote their dispersion directions, and the angular widths of each couple of colored arrows are the spectral extraction regions manually adjusted to cover the nuclear region of M51.}\label{fig:obsreg}
\end{figure}

Then \scriptname{\detokenize{merge_obs}} is invoked to reproject the observations and combine them together to create a merged event file and exposure-corrected images. 
Since we are only concerned about the distribution of diffuse hot gas in M51, point sources should be excluded in the image. 
The detection of point sources is conducted with \scriptname{wavdetect}. 
The list of the identified sources records the source locations and their sizes are represented by ellipses. 
The source list offers ingredients for \scriptname{roi} to remove emission in the source regions and for \scriptname{dmfilth} to refill these remaining holes based on local surrounding emission.
The generated image for diffuse hot gas (0.5--1.2~keV) in M51 is presented in \autoref{fig:obsreg} as the background intensity map.

\newpage
\subsection{Spatial Broadening Profiles}
\label{sect:prof}
We address here our method accounting for the broadening effect resulted from diffuse sources. 
Since \rgs s are slitless, lights from the whole field of view will be dispersed along the dispersion direction. 
The spatial extent of the radiating gas along this direction will lead to an effect of line broadening in the spectrum. 
The corresponding broadening effect can be quantified by $\Delta\lambda = 0.138\ \Delta\theta $ for the first-order spectrum, where $\Delta\lambda$ is the broadening width in \AA\ and $\Delta\theta$ is the source extent in arcmin. 
Convolution with spatial profiles of diffuse gas is necessary when dealing with the spectra of extended sources observed with \rgs s, and it is only possible to limit the spatial region of lights in the direction perpendicular to dispersion. 

\begin{figure}[htb!]\centering
\includegraphics[width=0.48\textwidth]{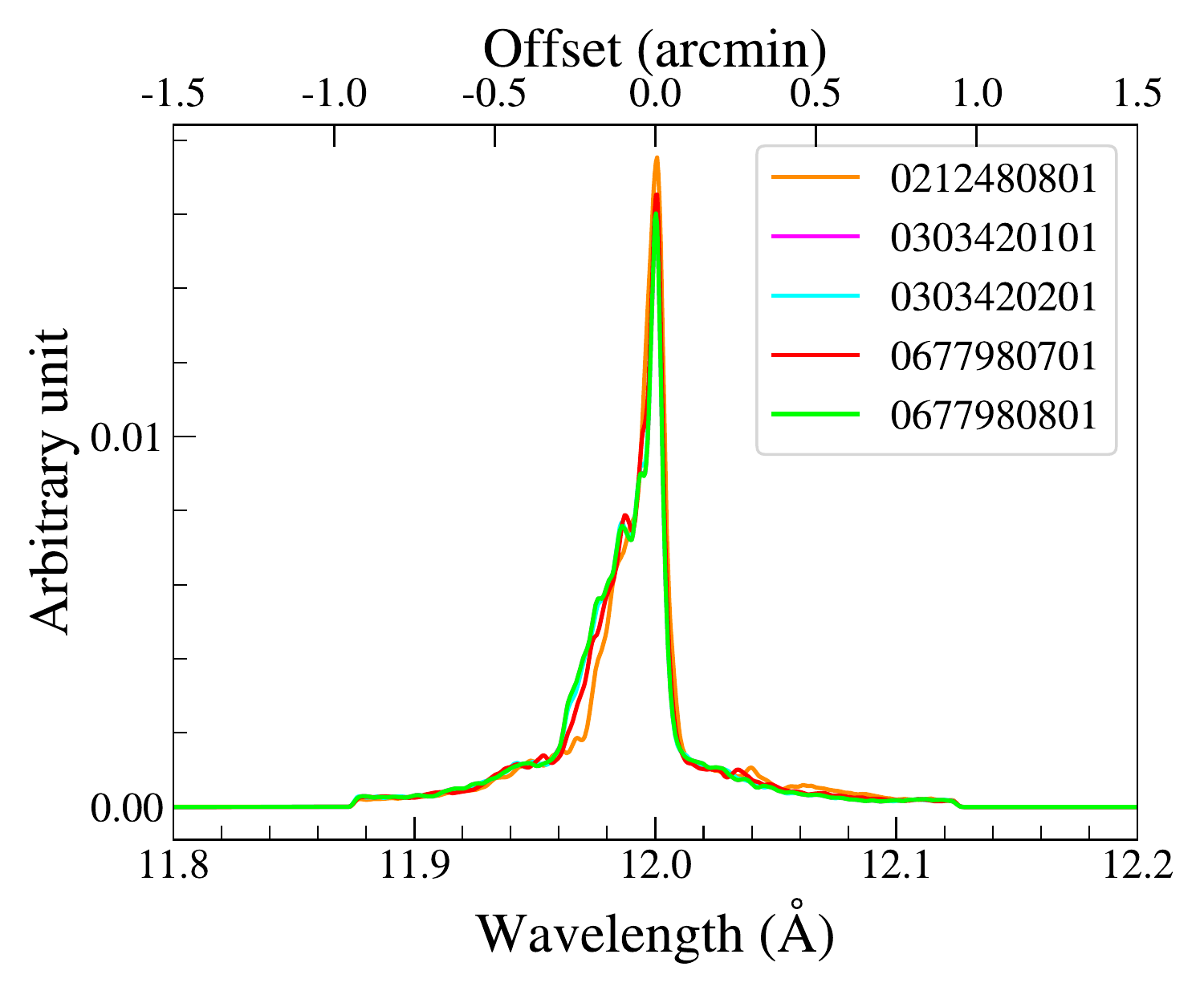}
\caption{Schematic spatial broadening profiles for the \xmm\ observations, generated from the \chandra\ diffuse gas image with energy of 0.5--1.2~keV. 
Each profile depicts the broadened line shape of an example Gaussian line centered at 12~\AA\ with a line width of 0.001~\AA . 
The spatial broadening effects of epochs of observations are quite similar because of their close PAs.}
\label{fig:broadening}
\end{figure}

The \chandra\ image is utilized to generate spatial broadening profiles for each \xmm\ observation. 
According to their different dispersion directions, each observation has its own region of extraction with one-arcmin cross-dispersion width, shown in \autoref{fig:obsreg} as parallel arrows. 
Photons outside the extraction regions are discarded, while inside the regions, images are compressed along the cross-dispersion direction to create the gas distribution profiles, which will be subsequently convolved with the model spectrum to account for the spatial broadening effect. 
The profiles are obtained with the help of the \modelname{rgsxsrc} model. 
During the process, the source image, boresight, PA, and size of the extraction region are needed. 
For consistency, the boresight is set to the X-ray brightest point and the size of the region is fixed at 1$^\prime$. 
\autoref{fig:broadening} illustrates the spatial broadening effects of the \xmm\ observations, by convolving their hot gas distribution profiles with a Gaussian line centered at 12~\AA\ with $\sigma = 0.001$~\AA .\par 

The variation of dispersion directions does not yield significant differences between the observations, as concluded from \autoref{fig:broadening}. 
We therefore adopt the observation of the longest exposure time (ObsID: 0303420101) as a representative, and use its profile to describe the spatial broadening effect for our model fitting. 
In the following fitting procedures, the profile should be convolved with the diffuse gas components.

\section{SPECTRAL ANALYSIS}
\label{sect:analysis}
The spectrum of the central region of M51 is highly line-dominated, as shown in \autoref{fig:trial}. 
The most conspicuous features include emission lines from H- and He-like ions of N, O, and Ne, in addition to a bunch of \FeXVII\ and \FeXVIII\ lines. 
We limit our analysis to the wavelength range of 7--28~\AA , where the signal-to-noise ratio is optimal.

We use pyXspec within {\it XSPEC} v12.10\footnote{https://heasarc.gsfc.nasa.gov/docs/xanadu/xspec/} to perform the spectral fitting procedure.
The Cash statistic is used, and the quoted errors refer to the 90\% confidence level for one free parameter.

\subsection{Fiducial Fit with a One-Temperature Model}
\label{sect:fiducial}
Given the complex physical conditions of the central region of M51, we start a fiducial fit by characterizing the diffuse hot gas with a single-temperature APEC model \citep{foster2012updated}.
The point source contribution is represented by a featureless power-law model. 
The foreground absorption for these two components is modeled by \modelname{tbabs} \citep{wilms2000absorption}, whose equivalent column density of hydrogen is fixed to the Galactic value\footnote{https://www.swift.ac.uk/analysis/nhtot/index.php} $N_\mathrm{H} = 1.93\times 10^{20}$~cm$^{-2}$ \citep{willingale2013calibration}. 

\begin{figure*}[htb!]\centering
  \includegraphics[width=0.95\textwidth]{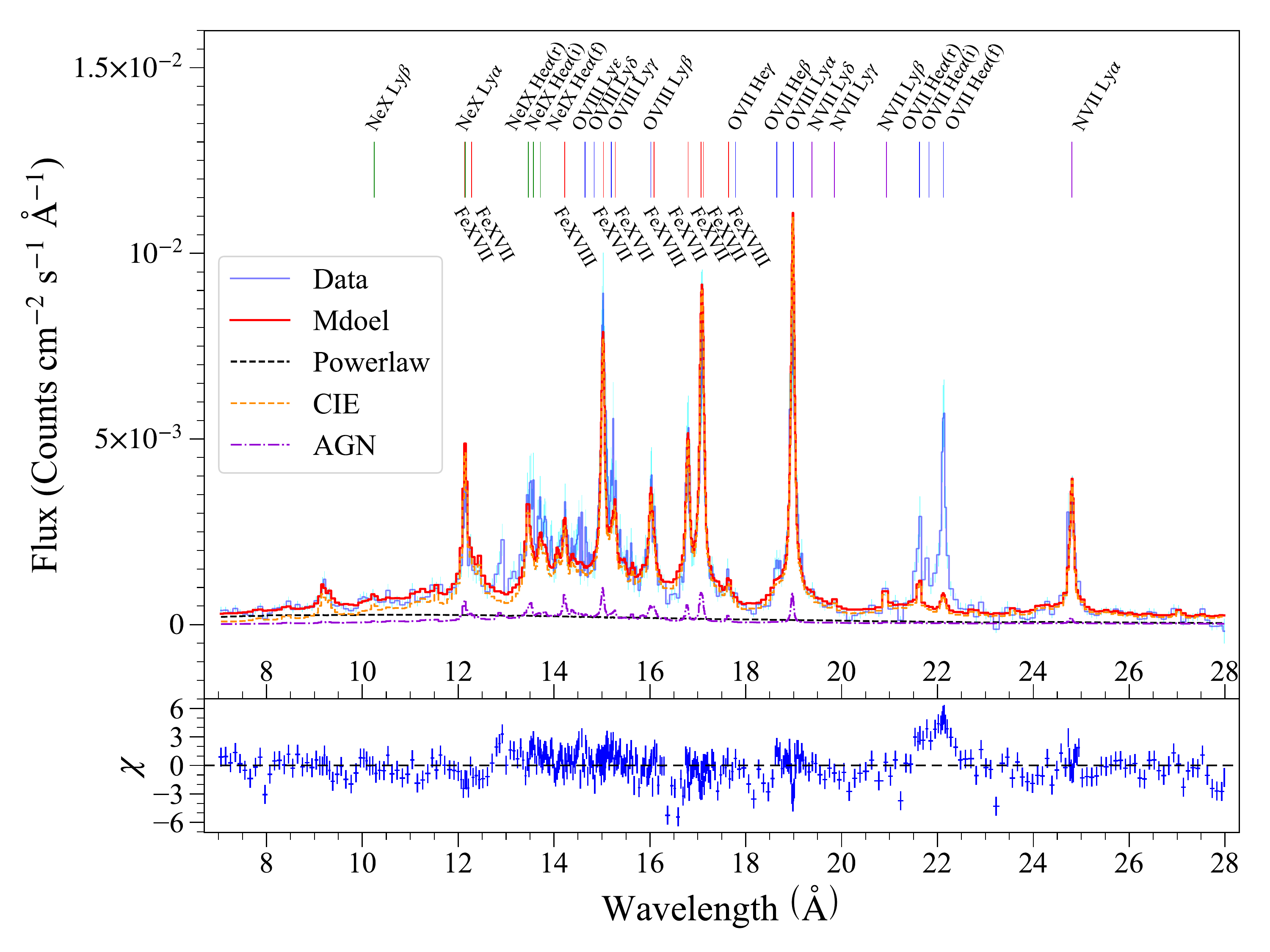}
  \caption{Fiducial fit of the spectrum, which is rebinned to reach a signal-to-noise threshold of 5 only for illustration. The temperature of the \modelname{VAPEC} is 0.48~keV, which can reproduce most of the observed spectral lines. 
The most incompatible feature between the observations and fitting model is the \OVII\ triplet at 21.5--22.5~\AA . 
Also shown as a purple dashed-dotted line is the AGN contribution to the spectrum, with parameters obtained from \citet{xu2016fluorescent}.
The vertical short lines indicate the redshifted wavelengths of emission lines in \autoref{tab:lines}.
For clarity, different elements are labeled with their unique colors, with N, O, Ne, and Fe in purple, blue, green, and red, respectively.}
\label{fig:trial}
\end{figure*}

The best-fit result is listed in \autoref{tab:par} (column 2) and plotted in \autoref{fig:trial}, illustrating that a single-temperature plasma of $\sim 0.49$~keV can model most of the line features.
The thermal component is in agreement with results by \citet{terashima2001chandra} and \citet{liu2015diffuse}, where a similar temperature was also reported when analyzing \chandra\ and \xmm\ data of the nucleus of M51, respectively.

Significant deviations remain at the wavelength ranges of 12.5--14.5~\AA\ and 21.5--22.5\AA .
The former range contains the \NeIX\ triplet and contribution from iron L-shell lines.
The deviation around the \NeIX\ He$\alpha$ triplet suggests the \NeIX\ forbidden line can be strong, though the \FeXIX\ (13.5 \AA) may also have some contribution.
The deviation around 13 \AA\ can be a \FeXX\ (12.85 \AA) line, accompanied by relatively weaker iron L-shell transitions spanning from 12.7 to 13.1 \AA .
In the latter range, the \OVII\ triplet is obviously not fitted, especially the strong forbidden line.
Additionally, the \OVIII\ Ly$\gamma$ line shows obvious excess compared to the model.

The deviations suggest the presence of unaccounted model components. 
The absence of an \OVII\ triplet in the model spectrum and the rich N abundance indicate that an additional component, behaving like a low-temperature thermal gas, is needed.
If the \FeXX\ line is real, a higher-temperature component may also be necessary.
In addition, the strong forbidden lines from the \NeIX\ and \OVII\ triplets do require the existence of some non-CIE process.

The fitting shapes of the \FeXVII\ lines and the \OVIII\ Ly$\alpha$ line are quite different.
It perspicuously suggests that the \FeXVII\ lines are narrower than the spatial broadening profile, while the \OVIII\ line is broader.
As a result, the \OVIII\ emission has a more extended distribution than that of the \FeXVII\ emission.

\begin{deluxetable}{lcc}[htb!]
\tablecaption{Parameters for different models \label{tab:par}}
\tablecolumns{3}
\tablehead{
\colhead{Parameters}				& \colhead{Single CIE}	& \colhead{CIE + CX} 
}
\startdata
$\eta_\mathrm{powerlaw}$\tablenotemark{*} ($\times 10^{-4}$)	& $0.48\pm 0.12$ 	& $0.76\pm 0.09$ \\
Photon index				& $0.66\pm 0.50$ 	& $1.46\pm 0.20$ \\ \hline
$\eta_\mathrm{vapec}$\tablenotemark{*} ($\times 10^{-4}$)			& $5.49\pm 0.54$ 	& $3.83\pm 0.58$ \\
$kT$ [keV]				& $0.49\pm 0.01$ 	& $0.59\pm 0.01$ \\ \hline
$\eta_\mathrm{vacx}$\tablenotemark{*} ($\times 10^{-4}$)			& --							& $1.10\pm 0.18$ \\
$v_{\rm relative}$	(\kmps)			& --			& 250 (fixed) \\ \hline
$\textrm{N} $ 			& $2.74\pm 0.34$ 	& $1.52\pm 0.24$ \\
$\textrm{O} $ 			& $0.46\pm 0.05$ 	& $0.39\pm 0.06$ \\
$\textrm{Ne}$ 			& $0.54\pm 0.06$	& $0.37\pm 0.06$ \\
$\textrm{Mg}$ 			& $0.44\pm 0.07$ 	& $0.42\pm 0.09$ \\
$\textrm{Fe}$ 			& $0.19\pm 0.02$ 	& $0.42\pm 0.06$ \\
Redshift 					& $0.0014(\pm3)$ 	& $0.0011(\pm3)$ \\
\hline
\multicolumn{1}{c}{$C$-statistic/d.o.f.} 		& 2710/2044 				& 2416/2043
\enddata
\tablecomments{The columns are: (1) free parameters in the models; (2) values of our fiducial fit (see \autoref{sect:fiducial}); and (3) values of our final result (see \autoref{sect:1TCX}). Five elements, N, O, Ne, Mg, and Fe, are allowed to vary, while other metal elements have solar abundances.}
\tablenotetext{*}{The normalization parameters of the \modelname{powerlaw} and \modelname{\it APEC} models are {\it XSPEC} defaults. For the power-law model, the `norm' has a physical meaning of $\rm photons\,keV^{-1}cm^{-2}s^{-1}$ at 1 keV. For the APEC models, the `norm' has a physical meaning of $\frac{10^{-14}}{4\pi[D_{\rm A}(1+z)^2]} \int n_{\rm e}n_{\rm H} dV$, where $D_{\rm A}$ is the angular diameter distance to the source (cm) and $z$ is the redshift, $n_{\rm e}$ and $n_{\rm H}$ are the electron and H densities (cm$^{-3}$), and $V$ is the volume. The `norm' of the \modelname{ACX} model is similar: $\frac{10^{-10}}{4\pi \left[ D_\mathrm{A} (1 + z) \right]^2} \int n_\mathrm{d} n_\mathrm{r} \mathrm{d} V_\mathrm{i}$, where $n_{\rm d}$ and $n_{\rm r}$ are the number densities of the donors and receivers in the CX process, and $V_\mathrm{i}$ is the volume of an interface layer where CX occurs.}
\label{tab:fit}
\end{deluxetable}

\subsection{Line Measurements and Line Ratios}
The fiducial fit also gives a reliable measurement of the continuum. 
Once determined, the continuum can be separated from individual emission lines, allowing the measurement of lines for spectroscopic diagnostics.

The line census is implemented by modeling all important emission lines simultaneously with Gaussian functions. 
The line centroids are fixed at their redshifted values deduced from the fiducial-fit redshift $z$. 
Each Gaussian is convolved with the profile obtained in \autoref{sect:prof} to account for the spatial broadening.
Since the thermal broadening is negligible compared with instrumental and spatial broadening, all the intrinsic line widths are set to 0.001\AA .  
The flux of each emission line is measured, as summarized in \autoref{tab:lines}. 

Line ratios calculated based on the individual line fluxes are presented in \autoref{tab:ratios}.
One can infer the temperature by the measured ratios, assuming that they originate from CIE plasma. 
However, the inferred temperatures from different groups of emission lines show great discrepancies. 
It seems that Fe line ratios favor a higher temperature than what O line ratios suggest. 
Some of the line ratios are not compatible with the CIE values between $10^6$ and $10^7$~K, and are denoted as \textit{N.A.} in the table. 
Nevertheless, the line ratios do imply non-CIE conditions of the nucleus of M51, which is consistent with our fiducial fit.

\begin{deluxetable}{cccc}[htb!]
\tablecaption{Line measurement of the central region of M51 \label{tab:lines}}
\tablecolumns{4}
\tablewidth{0pt}
\tablehead{
\colhead{Ion} & \colhead{$\lambda_\mathrm{rest}$ (\AA )} & \colhead{Transition \tablenotemark{a}} &
\colhead{Line Intensity \tablenotemark{b}}
}
\startdata
\NVII 	& 24.779 	& Ly$\alpha$ 						& 2.38$\pm 0.30$ \\
\NVII 	& 20.911	& Ly$\beta$ 							& 0.32$\pm 0.17$ \\
\NVII 	& 19.826 	& Ly$\gamma$ 					& 0.06$\pm 0.06$ \\
\NVII		& 19.361	& Ly$\delta$							& 0.15$\pm 0.11$ \\
\OVII 	& 21.602 	& He$\alpha$(r) 					& 1.21$\pm 0.27$ \\
\OVII 	& 21.800 	& He$\alpha$(i) 					& 0.71 ($=f_{22.100}/4.44$) \\
\OVII 	& 22.100	& He$\alpha$(f) 					& 3.16$\pm 0.32$ \\ 
\OVII		& 18.627	& He$\beta$							& 0.28$\pm 0.11$ \\
\OVII		& 17.768	& He$\gamma$						& 0.10$\pm 0.07$ \\
\OVIII 	& 18.973 	& Ly$\alpha$ 						& 3.98$\pm 0.23$ \\
\OVIII	& 16.003	& Ly$\beta$							& 0.93$\pm 0.15$ \\
\OVIII 	& 15.176	& Ly$\gamma$						& 0.85$\pm 0.16$ \\
\OVIII	& 14.821	& Ly$\delta$							& 0.05$\pm 0.05$ \\
\OVIII	& 14.634	& Ly$\epsilon$						& 0.63$\pm 0.10$ \\
\NeIX	& 13.448	& He$\alpha$(r) 					& 1.01$\pm 0.18$ \\
\NeIX	& 13.553	& He$\alpha$(i)					& 0.76$\pm 0.21$ \\
\NeIX	& 13.700	& He$\alpha$(f)					& 0.91$\pm 0.15$ \\
\NeX		& 12.132	& Ly$\alpha$						& 0.97$\pm 0.21$ \\
\NeX		& 10.239	& Ly$\beta$							& 0.24$\pm 0.08$ \\
\FeXVII 	& 17.096 	& 2p$^5$3s$^1 \rightarrow $ 2p$^6$ 	& 1.54$\pm 0.05$ \\
\FeXVII 	& 17.051 	& 2p$^5$3s$^1 \rightarrow $ 2p$^6$ 	& 1.71$\pm 0.06$ \\
\FeXVII 	& 16.780 	& 2p$^5$3s$^1 \rightarrow $ 2p$^6$ 	& 1.35$\pm 0.13$ \\
\FeXVII 	& 15.261 	& 2p$^5$3d$^1 \rightarrow $ 2p$^6$ 	& 0.94 ($=f_{15.014}/3$)\\
\FeXVII 	& 15.014 	& 2p$^5$3d$^1 \rightarrow $ 2p$^6$ 	& 2.82$\pm 0.15$ \\
\FeXVII	& 12.266	& 2p$^5$4d$^1 \rightarrow $ 2p$^6$	& 0.22 ($=f_{12.124}\times 0.92$) \\
\FeXVII	& 12.124	& 2p$^5$4d$^1 \rightarrow $ 2p$^6$	& 0.24$\pm 0.15$ \\
\FeXVIII	& 17.623	& 2p$^4$3p$^1 \rightarrow$ 2s$^1$2p$^6$	& 0.18$\pm 0.09$ \\
\FeXVIII & 16.071	& 2p$^4$3s$^1 \rightarrow $ 2p$^5$ 	& 0.69$\pm 0.11$ \\
\FeXVIII	& 14.208	& 2p$^4$3d$^1 \rightarrow$ 2p$^5$		& 1.11$\pm 0.13$ \\
\enddata
\tablenotetext{a}{In the notations of transitions of \FeXVII\ and \FeXVIII\ ions, the inner shell 1s$^2$2s$^2$ electron configuration is omitted for conciseness.}
\tablenotetext{b}{The line intensity is in units of $10^{-5}$ photons~s$^{-1}$~cm$^{-2}$. }
\end{deluxetable}

\begin{deluxetable}{lcc}[htb!]
\tablecaption{Line ratios used to determine temperatures \label{tab:ratios}}
\tablecolumns{3}
\tablewidth{0pt}
\tablehead{
\colhead{Lines} & \colhead{Ratio} & \colhead{Inferred Temperature \tablenotemark{a}}
}
\startdata
\OVIII\ Ly$\alpha$/\OVII\ He$\alpha$ 		& $0.78\pm 0.11$ 		& $\sim 3 \times 10^6$~K \\
\OVIII\ Ly$\beta$/Ly$\alpha$ 					& $0.23\pm 0.03$ 	& N.A. \\
\OVII\ He$\alpha$ G-ratio 							& $3.20\pm 0.78$ 	& N.A. \\
\NeX\ Ly$\alpha$/\NeIX\ He$\alpha$		& $0.36\pm 0.11$		& $\sim 4 \times 10^6$~K \\
\NeIX\ He$\alpha$ G-ratio        					& $1.65\pm0.46$    	&   N.A.  \\
$ \frac{\FeXVIII\ 16.071~\mathrm{\AA}}{\FeXVII\ 15.014~\mathrm{\AA}}$	& $0.24\pm 0.04$	& $\sim 6 \times 10^6$~K \\
[1em]
\FeXVII\ $ \frac{15.014~\mathrm{\AA }}{(17.051~\mathrm{\AA }+17.096~\mathrm{\AA })}$	& $0.87\pm 0.05$ 	& $> 10^7$~K \\
[1em]
\enddata
\tablenotetext{a}{The temperatures are inferred from CIE situation, where \textit{N.A.} means that the line ratio is out of the assumed range of $10^6$--$10^7$~K, indicating the associated lines are produced by either gas with much different temperatures or non-CIE processes.}
\end{deluxetable}

\section{SPECTRAL MODELING WITH CX}
\label{sect:fitting}
The complex nature of the observed spectrum calls for an alternative explanation beyond the commonly assumed CIE condition. 
Since M51 is likely to sustain multiphase outflows, CX can lead to a high forbidden-to-resonance ratio of He-like ions \citep[e.g.,][]{porquet2011helike, liu2012cxe}. 
The CX process describes a scenario in which highly ionized ions collide with neutral atoms/molecules, they steal electrons from the neutral species they encountered.
The transferred electrons tend to retain their potential energy; as a result, the excited ions emit photons in the X-ray regime as they cascade down.
In particular, for He-like ions after the CX reaction, during the electron transfer process, electrons are prone to accumulate in the $^{3}\mathrm{S}_1$ level, and thus strong forbidden lines are produced \citep[e.g.,][]{bodewits2007spectral, brown2009studies}. 

Another compelling reason to investigate the CX scenario is the fact that, with respect to the collisional process, its cross section (typically of the order of $\sim 3\times10^{-15}$~cm$^{-2}$) is much larger. Therefore, it is likely to contribute to the X-ray emission as long as both highly ionized and neutral species are interacting.
At the same time, the CX contribution behaves like a lower-temperature component due to the decreased ionization level.

By making use of the \modelname{ACX2} model developed by \citet{smith2012approximating}, we are able to account for the contribution from the CX process. 
With respect to the first version of this model, which applied empirical formulae when calculating reaction rates, the updated version now includes velocity-dependent actual cross sections, as well as more accurate energy-level-resolved data for electrons transferring to bare or hydrogenic ions. 
These new cross-section data are extracted from the modeling package \textit{Kronos} \citep{mullen2017}, and are also incorporated into another CX model in {\it SPEX} \citep{gu2016plasma}.
The final CX spectrum consists of the contribution from different combinations of neutral donors and recipient ions, each of which will be calculated separately.

\subsection{Fitting Results with the ACX Model}
\label{sect:1TCX}
With respect to the fiducial model in \autoref{sect:fiducial}, we investigate an additional CX component whose temperature, redshift, and metal abundances are tied to the thermal plasma. 
The interacting-velocity parameter is first allowed to vary and reports a best-fit value of $203 \pm 55$~\kmps\ that is mostly based on the \OVII\ G-ratio, while the derived acoustic velocity of the hot plasma is around 280~\kmps .
Comparably, the inferred radial velocity of the outflowing gas is $\sim300$~\kmps , based on the fitted redshift.
Therefore, we simply fix the interacting velocity at 250~\kmps .
As a result, we introduce only one new free parameter, the `norm' of the \modelname{ACX2} model.
Again, only abundances of N, O, Ne, Mg, and Fe are allowed to vary. 

During the fit, we apply the spatial broadening profile only to the \modelname{ACX2} model, but not to the thermal component \modelname{APEC} model.
This is because the \FeXVII\ line shape is narrower than the spatial profile, which means its emission has a more compact distribution.
In this case, it actually suggests a compact distribution of the thermal component, since the CX has little contribution to the \FeXVII\ lines.
Therefore, only the \modelname{ACX2} model uses the spatial profile, considering that the \OVIII\ line with a broader profile is significantly contributed by CX.

\begin{figure*}[htb!]\centering
  \includegraphics[width=0.95\textwidth]{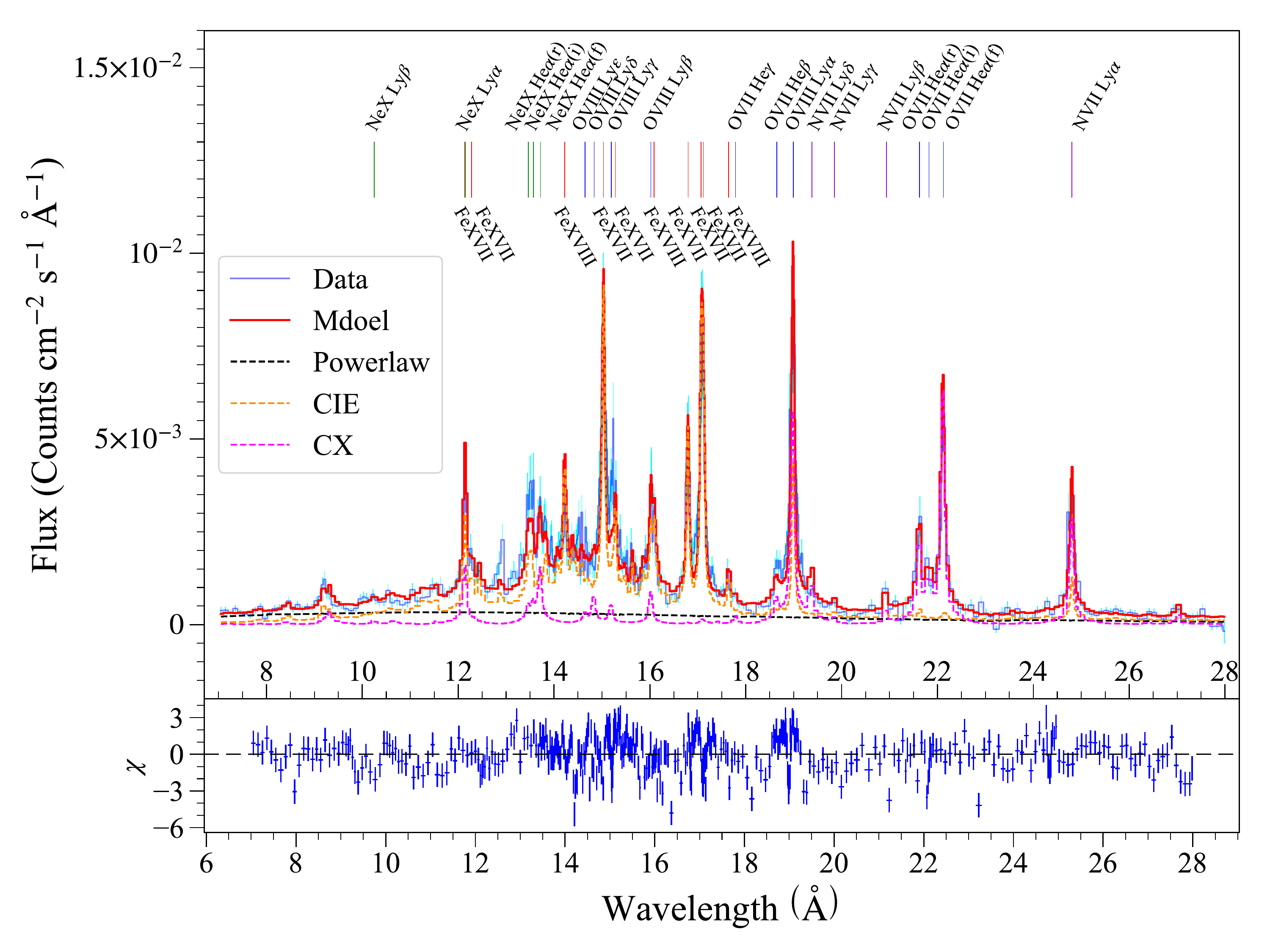}
\caption{Best-fit of a single plasma plus a CX component model; it is also as our final result. 
Both the temperature and the metal abundances of the two components are tied together. 
The CX component well explains the origin of the abnormal \OVII\ triplet forbidden line, and contributes to some other emission lines as well. 
In total the CX component accounts for $\sim 21\%$ of the continuum-subtracted flux.
The colors of the vertical lines are the same as those in \autoref{fig:trial}.}
\label{fig:1TCX}
\end{figure*}

The best-fit model is shown in \autoref{fig:1TCX} and it reduces the overall reduced-$C$-statistic from 1.3 to 1.2; the parameters are listed in \autoref{tab:par}. 
The temperature increases from 0.49 to 0.59 keV.
The overall metal abundances slightly decrease except for Fe, yet they are consistent with a typical interstellar level. 
Significantly dropped is the N abundance, which is overabundant in the single-CIE scenario, but now reduces to roughly one solar value.
Overall, the fit considerably reduces the residuals of the fiducial fit.
In total, the CX contribution accounts for $\sim 21\%$ of the total diffuse X-ray emission in the range of 7--28~\AA .

The CX component produces strong \OVIII\ and \OVII\ forbidden lines, the latter of which solve the problem of the G-ratio.
The most special and unique features of the CX radiation are the higher-order Lyman lines such as the \NVII\ Ly$\delta$ and \OVIII Ly$\gamma$ or the higher-order K-shell lines such as the \OVII\ K$\gamma$ line.
The \OVIII\ Ly$\delta$ is kind of weak but the Ly$\epsilon$ seems to be stronger in the spectrum.
The higher-order of Sulfur Lyman lines \citep{gu2015novel, hitomi2017constraints} in the Perseus cluster are deemed to be the signature of CX.
Our analysis shows that the CX scenario, suggested by similar strong high order Lyman lines, satisfactorily explains the physical conditions of the nuclear region of M51.
It proves that the inclusion of a CX component tenders a possible scenario for the physical circumstances of M51's center.

\subsection{Location of the CX Emission}
\begin{figure*}[htb!]\centering
  \includegraphics[width=0.95\textwidth]{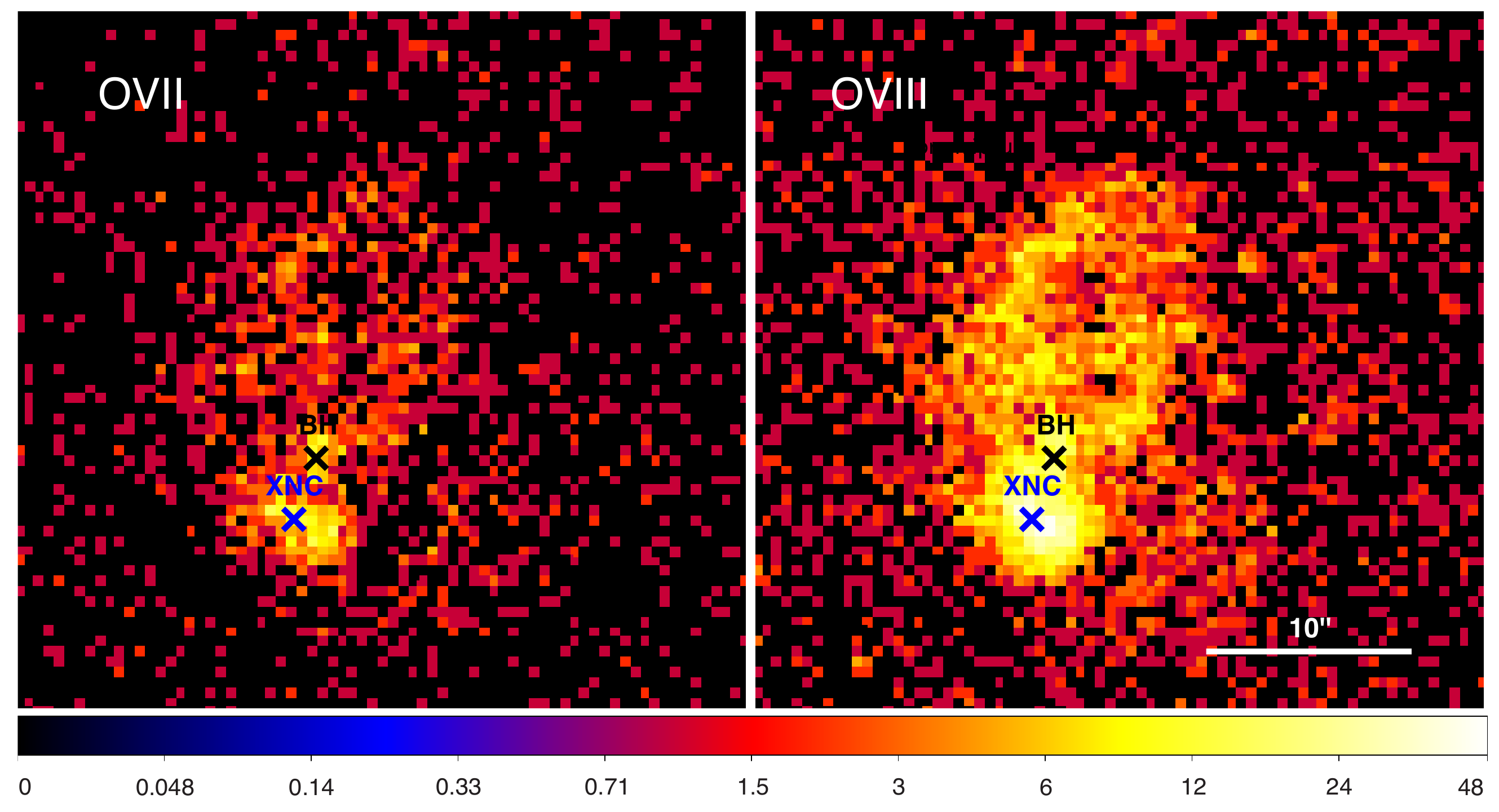}
  \caption{Monochromatic maps of \OVII\ and \OVIII\ from \chandra\ observations in logarithmic scale.
  The black `x' mark is the position of the SMBH, and the blue one is the brightest point in the soft X-ray band image, which lies in the extranuclear cloud. Left: \OVII\ (0.551--0.585~keV); right: \OVIII\ (0.620--0.689~keV).}
\label{fig:linemap}
\end{figure*}

To locate the region where the CX takes place would be quite useful.
We extract monochromatic maps from Chandra data in the bands for the \OVIII\ line (0.620--0.689~keV) and the \OVII\ line (0.551–0.585 keV), as shown in \autoref{fig:linemap}.
Since the continuum emission in these bands is negligible, the two maps roughly represent the line emission here, though their photons may be slightly blended due to the limited energy resolution of \chandra\ CCD spectra.
More than 75\% of the emission relates to the structure of jet-driven outflows, as will be detailed in the next section.
As a result, the prominent \OVII\ forbidden line must be emitted from these regions.

On the other hand, the \OVII\ forbidden line should be more noticeable along the southern outflow.
Using the \xmm /\rgs\ data, \citet{liu2015diffuse} has reported that the peak of the \OVII\ forbidden line is about $+10''$ offset from that of other lines along the cross-dispersion direction.
This result is naturally consistent with the southern location, since the jet-driven outflow in the north is undergoing a dissipative phase in which less CX emission being produced is also reasonable.
Based on these results, we further discuss about the origin of the diffuse soft X-ray emission.

\section{DISCUSSION}
\label{sect:discussion}
\subsection{Case of AGN Jet-driven Outflow}
In the nuclear region of M51, the hot gas might originate from shocks caused by the radio jet \citep{bradley2004physical, maddox2007study}, whose effects are deemed to be confined mainly within the central $\sim 1$~kpc region.
The hot gas has a bipolar structure, and following previous studies it is referred as a northern loop plus an extranuclear cloud (XNC) in the southeast, which is also presented in radio observations \citep{crane1992radio, terashima2001chandra, bradley2004physical}.
In particular, the XNC is brighter in soft X-ray compared with the Northern Loop, and the asymmetrical morphology indicates possible one-sided jet ejection \citep{crane1992radio, maddox2007study}, or distinct ISM conditions of north and south sides.

Since M51 is a star-forming galaxy that may contain stellar-driven nuclear wind, it is nontrivial to further separate the contribution of the XNC from the wind. 
A potential revelation is provided by the morphology seen in the \chandra\ image (see \autoref{fig:obsreg}), from which we shrink the extraction region to a smaller region with a diameter of $0.5'$. 
The diameter is chosen to contain the northern loop and XNC, leaving the outer region filled with more diffuse gas. 
Then, the resultant spectrum shows similar relative intensities of spectral lines, only reducing the count rate by about a quarter. 
This result also inspires us to attribute the most X-ray emission to the AGN jet-driven outflow.

The inclination angle of the radio jet with respect to the disk is 15$^\circ$ \citep{cecil1988kinematics}, indicating that the jet pierces into the dense disk plane, interacts with the  ISM, and forms the structures seen in radio and X-ray \citep[see Fig.\,14 in][]{querejeta2016agn}. 
The shock-heated hot gas in the XNC shows the brightest surface X-ray emission.
It would then move toward the southeast along the jet direction, and do the expansion simultaneously.
Based on the fitted redshift of \modelname{APEC}, the radial velocity of the hot gas along the line of sight is about 300 \kmps , suggesting a probably larger outflowing velocity of the jet-driven outflow.
The ionized outflow further interacts with the ambient neutral gas and produces CX emission.
As a result, the CX occurs roughly at the southeast side of the XNC, in agreement with the location we deduce previously.

In this scenario, only the XNC is of concern since it dominates the emission in the extracted region.
It spans about $5''$ in the sky, corresponding to a diameter of $\sim 200$~pc. 
Therefore, assuming a sphere geometry, the hot gas-filling volume $V = 1.23 \times 10^{62}$~cm$^{3}$, and the surface area $S = 1.20 \times 10^{42}$~cm$^{2}$. 
From the definition of the \modelname{APEC} `norm' (as in \autoref{tab:par}), the hydrogen density of the hot outflow is $n_\mathrm{H} = 1.51$~cm$^{-3}$.
Then the total mass of the hot gas is $M_\mathrm{hot} = 1.56 \times 10^5~M_\odot$, and the injected energy is $\sim10^{53}$ erg.

In the definition of the \modelname{ACX2} `norm,' the $V_\mathrm{i}$ is the volume of an interface layer where CX occurs, which can be expressed as the product of the interface area $A$ between the hot and cold gas and the penetration depth $l$ of ions.
Thus, the terms $n_\mathrm{d}$ and $V_\mathrm{i}$ in `norm' are physically degenerated.
When the density of cold gas increases, given the amount of incident ions $n_\mathrm{r}$, the corresponding effective interaction volume may decrease because the ions can now penetrate a thinner layer of the cold gas clouds and then become neutralized because of successive CX reactions.
Considering the depth of penetration $l$ being the mean-free path length of hot ions, it can be derived from $l = \left( \sigma n_\mathrm{d} \right)^{-1}$. 
Note that the interacting velocity is taken into account outside the `norm' calculation.
Accordingly, the `norm' of \modelname{ACX2} can then be expressed as
\begin{equation}
  \eta_\mathrm{vacx} = \frac{10^{-10}}{4\pi \left[ D_\mathrm{A} (1 + z) \right]^2} \int \frac{n_\mathrm{r}}{\sigma} \mathrm{d} A
  \label{eq:vacx_norm_explicit}
\end{equation}

Taking the distance of M51 as $D = 2.64\times 10^{25}$~cm \citep{mcquinn2016distance}, and adopting a typical value of the CX cross section $\sigma \sim 3\times 10^{-15}$~cm$^{-2}$, we can get the effective interface area $A = 1.92 \times 10^{43}$~cm$^2$.
The effective area for CX process $A$ is roughly an order of magnitude larger than the geometrical surface area $S$, similar to that reported in M82 superwind \citep{zhang2014m82}.
It can be interpreted by turbulences that resulted from Kelvin-Helmholtz and Rayleigh-Taylor instabilities between the multiphase layers, until thermal conduction dominates at tiny scales.
But quantitative estimation of how turbulence effectively enlarges the interaction area still relies on future studies.

\subsection{Case of Stellar Feedback Outflow}
It has been reported that the central region of M51 has undergone relatively active SF \citep[e.g.,][]{calzetti2005star, schinnerer2013paws}, and young stellar components are also resolved in this region \citep[e.g.,][]{scoville2001highmass, maddox2007study}.
It is reasonable to consider this extreme scenario: the nuclear outflow is completely triggered by central SF and propagates outward to interact with ambient neutral gas to produce CX emission.
Despite the studies for the XNC, the bright soft X-ray emission could be where the superwind is well confined in the dense disk plane.
When the superwind breaks through the disk plane, it expands outward and shows more diffuse X-ray emission.

It is important to establish the geometrical structure of the superwind before we make further estimations. 
Commonly assumed is a biconical structure of the outflow similar to that presented in the prototype starburst galaxy M82 \citep{melioli2013evolution}, which is also shown in recent simulations \citep[e.g.][]{sarkar2015}. 
However, the face-on orientation of M51 hampers us from identifying the geometry of the galactic superwind, even its presence. 
In spite of the small portion of outflowing gas that resides in the uncertain opening angle at the edge of this region, the geometric structure can be simplified as a cylinder whose bottom is the extraction region and whose height is the outermost radius $R_\mathrm{outflow}$ that the outflowing gas reaches.
 
While the bottom area $S_\mathrm{bottom} = \pi \left(D_\mathrm{A} \times 1~\mathrm{arcmin}\right)^2 = 1.85 \times 10^{44}$~cm$^2$ can be readily calculated, the estimation of $R_\mathrm{outflow}$ still demands reliable evidence. 
Despite the difficulty of accurately determining the value of $R_\mathrm{outflow}$ from observations, rough estimations can be made by comparing with edge-on analogs or suggested values from numerical simulations. 
Two representative values of $R_\mathrm{outflow}$ are taken: 3~kpc as presented in the case of M82 \citep{melioli2013evolution}, and 10~kpc from a simulation \citep{sarkar2015} with an energy input rate of $L \sim 10^{41}$~erg~s$^{-1}$ assumed. 
By substituting the `norm' values in \modelname{ACX}, one can obtain
\begin{equation*}
\left \{
\begin{aligned}
  n_\mathrm{H} &= 1.28 \times 10^{-2}~\mathrm{cm}^{-3}, & R_\mathrm{outflow} &= 3~\mathrm{kpc}\\
  n_\mathrm{H} &= 7.00 \times 10^{-3}~\mathrm{cm}^{-3}, & R_\mathrm{outflow} &= 10~\mathrm{kpc}
\end{aligned}
\right.
\end{equation*}
The total mass of hot plasma is then $M_\mathrm{hot} \simeq m_\mathrm{H} n_\mathrm{H} V = 1.84 \times 10^7~M_\odot$ for $R_\mathrm{outflow} = 3~\mathrm{kpc}$, and $M_\mathrm{hot} = 3.36\times 10^7~M_\odot$ for $R_\mathrm{outflow} = 10~\mathrm{kpc}$.

Again we can infer the effective interface area of CX process:
\begin{equation*}
\left \{
\begin{aligned}
  A &= 2.26 \times 10^{45}~\mathrm{cm}^2, & R_\mathrm{outflow} &= 3~\mathrm{kpc}\\
  A &= 4.13 \times 10^{45}~\mathrm{cm}^2, & R_\mathrm{outflow} &= 10~\mathrm{kpc}
\end{aligned}
\right.
\end{equation*}
which is also an order of magnitude greater than the geometrical surface area $S_\mathrm{bottom}$, implying that the two phases of gas are well mixing to increase the interacting area.

This scenario suffers the setback that the \chandra\ images are not centralized, while the superwind is commonly assumed to be perpendicular to the disk and have a large opening angle. 
The current SF in the central region of M51 is not that severe.
The reported SFR in the central region ranges from $0.24~M_\odot$~yr$^{-1}$ \citep{rampadarath2015high} to $\sim 1~M_\odot$~yr$^{-1}$ \citep[based on some measurements of spatially resolved SFR densities, see][]{kennicutt2007star, leroy2017cloudscale}.
However, hundreds of \HII\ regions and young stellar components identified within the central region \citep{scoville2001highmass} suggest potential feedback from previous violent SF.
The hot gas outflow may now come into a tenuous state, as hinted by the somewhat centralized diffuse emission within the radius between $0.5^\prime$ and 1$^\prime$ to the central black hole.

As a consequence, both the mechanisms probably contribute to the outflow.
The jet-driven outflow dominates in the inner region as revealed by the bright X-ray emission, while the outer region fully fills with the tenuous hot gas from the SF-driven outflow.
These ionized outflows all interact with the ambient neutral gas, undergoing the CX process.
Of course, hot gas produced by jet-deduced shock and by stellar feedback do not need to share similar properties.
The emission lines in the RGS spectrum mainly originate from the jet-driven outflow, but the unfitted broad wings from the \OVIII\ Ly$\alpha$ or the \NeIX\ He$\alpha$ suggest the emission from more diffused stellar feedback outflow.

\subsection{Influence on the cold gas accumulation}
Previous observations reveal that in comparison to its spiral arms, the central region of M51 is not a great reservoir of neutral gas, either for \HI\ \citep{walter2008things} or for H$_2$ traced by CO \citep{schinnerer2013paws}. 
\citet{schuster2007complete} estimated the radial distributions of both species, finding that within the radius of 1~kpc, surface densities of \HI\ and H$_2$ are $\sim 6~M_\odot$~pc$^{-2}$ and $\sim 40~M_\odot$~pc$^{-2}$, respectively.
It is interesting to check how fast the cold gas would be destroyed due to the interaction with hot gas.

Obtained from the fitting result, the ion incident rate of CX is $f_\mathrm{r} = \int n_\mathrm{r} v \mathrm{d}A \simeq ~7.23\times 10^{50}$~s$^{-1}$, with a 250 \kmps\ interacting velocity.
If that each ion consumes a neutral particle, the total cold gas consumed would be $\lesssim 19 ~M_\odot$~yr$^{-1}$.
Even for the case of jet-driven outflow within a radius of about 200 pc, the existing cold gas is sufficient for the CX process up to 0.1 million years.
It seems that the CX has no vital influence on the cold gas accumulation.

\subsection{Possibility of AGN Photoionization}
AGN photoionization is also a possible mechanism producing the strong \OVII\ forbidden line, as hydrogen-like ions are recombined into helium-like ions.
Based on the \chandra\ and \nustar\ observations, \citet{xu2016fluorescent} and \citet{brightman2018long} investigated the intrinsic luminosity of M51, claiming an order of $\sim 10^{40}$~erg~s$^{-1}$ in 2--10~keV for the AGN. 
It is only about 10\% of the luminosity of the whole galaxy in the same energy band ($\sim 10^{41}$~erg~s$^{-1}$; \citep{lutz2004relation}).
Moreover, using the inferred parameters of the thermal component from \citet{xu2016fluorescent}, we find that spectral contribution of the M51 nucleus (purple dasded-dotted line in \autoref{fig:trial}) is negligible even in such a $1^\prime$ region.

The photoionized gas requires an ionization parameter $\xi=L/nr^2>5$ to be responsible for the strong \OVII\ forbidden line.
For the XNC that is $\sim$200 pc away from the black hole and has a density of 1.51 cm$^{-3}$, the ionization parameter $\log (\xi)=-0.76$ is too small.
This level of photon injection by the current AGN does not seem at all sufficient. 
Additionally, \citet{liu2015diffuse} also suggested that the intensities of Fe L-shell lines are too high to be consistent with a photoionization plasma. 

Although it seems the current activity of the nucleus of M51 cannot be responsible for the observed spectrum, a past AGN outbreak remains as a candidate, since it is not completely excluded by the higher-order Lyman series of N and O listed in \autoref{tab:lines}.
If the AGN of M51 had a much higher luminosity in its recent history, a recombining plasmas could possibly produce features similar to those of our measurement.
Plausible as this is, such a scenario is less convincing because of the lack of the M51 AGN history, as well as the absence of an ionization gradient along the northern outflow structure that elongates out to about 500~pc (\autoref{fig:linemap}).

\subsection{Possibility of Resonance Scattering}
Resonant scattering is another process in favor of a high forbidden-to-resonance ratio, for it reduces the intensity of the resonance line. 
In order to quantify the effect of resonant scattering in M51, we calculate the optical depth of the \OVII\ resonance line. 
The optical depth can be estimated by 
\begin{equation}
\label{eq:tau}
\tau = \int_0^{R_\mathrm{max}} n_\mathrm{i} \sigma \mathrm{d}r,
\end{equation}
where $R_\mathrm{max}$ is outermost radius of the outflow, and the ion density $n_\mathrm{i}$ should be calculated correspondingly. 
The line-center cross section $\sigma$ can be written as \citep[e.g.,][]{rybicki1986}
\begin{equation}
\sigma_{\nu_0} = \frac{\sqrt{\pi}e^2}{m_\mathrm{i} c \Delta \nu_\mathrm{D}}f_{lu},
\label{eq:xsect}
\end{equation}
where $\Delta \nu_\mathrm{D} = \frac{\nu_0}{c} \sqrt{\frac{2k_\mathrm{B}T}{m_\mathrm{i}}}$ is the Doppler width, $\nu_0$ is the frequency of the line center, $e$ is the elementary charge, $m_\mathrm{i}$ is the ion mass, $c$ is the speed of light, and $f_{lu}$ is the oscillation strength of the transition from a lower to an upper level. And the ion density $n_\mathrm{i}$ can be obtained from $n_\mathrm{i} = n_\mathrm{H}Z f_\mathrm{i}$, where $Z$ is the metal particle portion relative to the hydrogen number density and $f_\mathrm{i}$ is the ionization fraction. 

We adopt the oxygen particle portion of $Z=4.9\times 10^{-4}$ from \citet{wilms2000absorption} and oxygen abundance of 0.5 solar value, $f_{\OVII} \simeq 2.7\times 10^{-3}$ at $T=0.5$~keV from the single-temperature fitting, and the oscillation strength $f_{lu} = 0.72$ for the oxygen resonance line from atomic database AtomDB\footnote{http://www.atomdb.org/} \citep{foster2012updated}. 
In combination with the values derived previously, for the \OVII\ resonance line we find its line-center optical depth is sufficiently small ($\tau \sim 10^{-6}$) for different sets of assumptions in \autoref{sect:discussion}.

In a region small as the central part of M51 where gas with fast motions tends to give rise to broader line width, it is hard to reach significant optical depths for resonant scattering to matter, which affects line wings less seriously than the line centroid.
Thus, the effect of resonant scattering can be ignored in M51.

\section{SUMMARY}
\label{sect:summary}
We investigate the spectrum of the central region of M51 observed with \xmm/\rgs , with a total effective exposure of $\sim 373$ ks. 
The high spectral resolution enables us to study the underlying processes of the hot plasma in the soft X-ray regime based on line diagnostics. 
We present a detailed spectral analysis and fit the spectrum with the up-to-date CX model.
The main results and conclusions are presented as follows:
\begin{enumerate}

\item 
In the \rgs\ spectrum, the \OVIII\ G-ratio, the \NVII , and the \OVIII\ Ly$\gamma$/Ly$\beta$ ratio are abnormally high when compared with predictions of a CIE plasma, suggesting evidence of CX emission. 
Using the newest CX model \modelname{ACX2} plus the CIE model \modelname{APEC}, we fitted the entire \rgs\ spectrum, and found that CX contributes to about one-fifth of the gas emission in the wavelength range of 7--28~\AA . 
The temperature of the hot gas is 0.59 keV and the metal abundances of O, Ne, Mg, and Fe are sub-solar. 
The abundance of N is slightly higher than the solar value. 

\item 
Our result favors the scenario where the outflow, containing a mass of about $10^5 M_{\odot}$, is driven by the jet of the AGN, injecting an energy of about $10^{53}$ erg.
The outflow moves outward to the southeast side, and interacts with the ambient neutral gas to produce significant CX emission.
The effective interface area of the hot and cold gas is about $10^{43}$ cm$^2$, roughly one order of magnitude larger than the surface area of the outflow.

\item 
The stellar feedback outflow in the central region may also contribute to the X-ray emission, through both the thermal and the CX emission.
However, the CX process can only consume the cold gas for about 19 $M_{\odot} \mathrm{yr}^{-1}$, which does not significantly impact the accumulation of cold gas.

\item 
The current low luminosity of the AGN in M51 disfavors an ongoing photoionization scenario as a possible explanation for the presence of strong forbidden lines. 
But past AGN activity remains plausible.
  
\item
The resonant scattering is unlikely due to the small optical depth, especially when considering the rapid motion of hot gas in the central region of the galaxy.

\end{enumerate}
\par 

\acknowledgments
We acknowledge the anonymous referee for the constructive suggestions. 
We thank Q.~Daniel~Wang, Randall~Smith, J\"orn~Wilms, and Jiren~Liu for insightful and helpful discussions, thank Adam~Foster for support with the \modelname{ACX2} model, and thank Wei~Sun for technical help with the \chandra\ data reduction. 
S.N.~Zhang acknowledges the support from NSFC grant 11573070.
L.~Ji acknowledges the support from NSFC grant U1531248.

\bibliography{M51}

\end{document}